\documentstyle[12pt,epsfig]{article}
\newcommand{\be}{\begin{equation}}
\newcommand{\ee}{\end{equation}}
\newcommand{\bea}{\begin{eqnarray}}
\newcommand{\eea}{\end{eqnarray}}

\def\cR{{\cal R}}

\begin{document}
\title{
  \begin{flushright} \begin{small}
  DTP--MSU/00-05 \\ hep-th/0005099
  \end{small} \end{flushright}
\vspace{2.cm}
{\bf Solitons and black holes in non-Abelian Einstein-Born-Infeld theory}}

\bigskip
\author{ V.V. Dyadichev\thanks{Email: rkf@mail.ru}
      and  D.V. Gal'tsov\thanks{Email: galtsov@grg.phys.msu.su} \\ \\
{\it Department of Theoretical Physics,}   \\
       {\it  Moscow State University, 119899, Moscow, Russia}
}
\date{May 6, 2000}
\maketitle

\begin{abstract}
Recently it was shown that the Born--Infeld--type modification of the
quadratic Yang--Mills action gives rise to classical particle-like solutions
in the flat space which have a striking similarity with the Bartnik-McKinnon
solutions known in the gravity coupled Yang-Mills theory. We show that both
families are continuously related within the framework of the
Einstein-Born-Infeld theory through interpolating sequences of parameters.
We also investigate an internal structure of the associated black holes. It
is found that the Born--Infeld non--linearity leads to a drastic modification
of the black hole interior typical for the usual Yang-Mills theory. In the
latter case a generic solution exhibits violent metric oscillations
near the singularity. In the Born-Infeld case a generic interior solution is
smooth, the metric has the standard Schwarzschild type singularity, and
we did not observe internal horizons. Such smoothing of the 'violent' EYM
singularity may be interpreted as a result of quantum effects.

\end{abstract}
\newpage

Physical significance of the particle-like solutions to the
Einstein-Yang-Mills (EYM) field equations found by Bartnik and McKinnon (BK)
\cite{BaMc88} (for more details see a review paper \cite{VoGa98}) as well
as their possible role in the string-inspired models remains rather obscure.
These 'particles' where found to have a sphaleronic nature \cite{GaVo91}
and they could be responsible for fermion number violating effects. However,
it is not clear whether these purely classical solutions
can survive the embedding into some quantum framework. Another related puzzle
is the singularity structure of the associated black holes \cite{VoGa89,
Ku90,Bi90}. It was shown that the metric inside the classical EYM black holes
exhibits violent oscillations near the singularity, which go far beyond the
classical bounds \cite{DoGaZo96,GaDoZo97,BrLaMa98}. Presumably these
oscillations must be regularized in the quantum theory, but no relevant
explanation was suggested so far.

To probe the effect of quantum corrections to the Yang-Mills lagrangian
within the string/M theory framework it is tempting to utilize the Born-Infeld
modification of the Yang-Mills action describing the low energy
dynamics of D-branes \cite{Po98}. Classical solutions to both the Abelian
and non-Abelian Born-Infeld theories received recently much attention.
Although there are certain complications in the definition of the trace
over the gauge group generators in the Born-Infeld action
\cite{Ts97,GaGoTo98,BrPe98}, it is believed that the simplified 'square root'
form of the lagrangian gives correct (at least qualitatively) predictions
concerning solitons. Magnetic monopoles were shown to persist in such
a theory \cite{GrMoSc99,Pa99,Gr99}.
A new type of a particle-like solution in the non-Abelian Born-Infeld
model was obtained by Gal'tsov and Kerner (GK) \cite{GaKe99}.
It was shown that this theory gives rise to flat space classical
glueballs which have a striking similarity with the BK solutions.

More close relationship between these two particle-like configurations
becomes clear within the Einstein-Born-Infeld (EBI) theory. As was
shown recently by Wirschins, Sood and Kunz \cite{WiSoKu00}, the GK solutions
survive when gravity is added, in which case the corresponding black holes
also come into play. There is
a substantial difference between the above two families which has to be
clarified, however. Both the BK and GK particles form
the same kind of discrete sequences resulting from the 'quantization'
of the parameter entering the boundary conditions near the origin. But
contrary to the BK case for which this parameter sequence is convergent
to a limiting value, in the GK case the corresponding sequence is divergent.
So it is necessary to check more carefully whether the both families are
continuously related indeed.

Here we investigate the non-Abelian
Einstein-Born-Infeld solitons and black holes in more detail. We show that
within this framework one has a unique family of the regular particle-like
solutions smoothly interpolating between the GK and BK solutions while the
gravitational coupling constant varies from zero to a large value in units
of the Born-Infeld 'critical field' parameter. We also investigate the
interior structure of the non-Abelian EBIYM black holes and find that
the problem of violent metric oscillations here is resolved indeed.

Recall that the
static spherically symmetric EYM field equations admit three smooth branches
of local solutions  near the singularity exhibiting the Schwarzschild,
Reissner-Nordstr\"om and the imaginary-charge Reissner-Nordstr\"om type
behavior. Neither of these three, however, has a sufficient number of free
parameters to be generic. Therefore when one is moving from the event horizon
into the black hole interior one can meet smooth local solutions near the
singularity at best for discrete values of the black hole mass
\cite{DoGaZo96}. The generic
EYM black hole interior looks very differently from other explicitly known
cases. When the singularity is approached the metric exhibits oscillations
with an infinitely growing amplitude and an infinitely decreasing period.
Clearly, the classical bounds are exceeded  after a few oscillation cycles.
In the non-Abelian EBI theory we can also find several local solution
branches near the singularity, but the situation is drastically different.
The local solution  which has a sufficient number of parameters to be generic
has a perfectly smooth Schwarzschild type behavior. Therefore the problem
of oscillations is resolved in a natural way.

As in \cite{GrMoSc99,GaKe99,WiSoKu00} we assume the 'square root' form of the
non--Abelian Born--Infeld action which, as we believe, describes well enough
particle-like solutions being at the same time much simpler than
the favored by strings symmetrized trace \cite{Ts97} action.
Thus the EBIYM action is chosen as
\be \label{S}
S=\frac{1}{16\pi G}\int\;\left( -R+4G\beta^2(1-\cR)\right)\;d^4x,
\ee
where
\be
\cR=\sqrt{1+\frac{1}{2\beta^2}
F^a_{\mu\nu}F_a^{\mu\nu}
-\frac{1}{16\beta^4}(F^a_{\mu\nu}{\tilde F}_a^{\mu\nu})^2}.
\ee
Without loss of generality the dimensionless gauge coupling constant
(in the units $\hbar=c=1$) will be set to unity, so we are left with two
(dimensionfull) parameters: the BI 'critical field'
$\beta$ of dimension $L^{-2}$, and the Newton constant $G$ of dimension
$L^{2}$ (Planck's length). From these one can form a dimensionless
constant
\be
g=G\beta,
\ee
which is the only substantial parameter of the theory \cite{WiSoKu00}.
The decoupling of gravity corresponds to the limit $g\to 0$, while the
$g\to \infty$ limit (after rescaling) gives the EYM theory.

We consider the case of the $SU(2)$ gauge group assuming for the YM field
a usual spherically symmetric static purely magnetic ansatz
\be    \label{Aa}
A=(1-w(r))\left(T_\theta \sin\theta d\varphi -T_\varphi d\theta\right) ,
\ee
where a rotated basis for the gauge group generators is used.
The spacetime metric is parameterized as follows:
\be
ds^2=N\sigma^2 dt^2-\frac{dr^2}{N}-
r^2\left(d\theta^2+\sin^2\theta d\varphi^2\right).
\ee
The equations of motion after a rescaling of the radial coordinate
$(\beta)^{1/2}r\to r$ (making $r$ dimensionless) take the form
of two coupled equations for $N$ and $w$:
\be \label{eqw}
\left(\frac{Nw'}{\cR}\right)'= \frac{w(w^2-1)}{r^2\cR}-\frac{2gw'^3 N}{\cR^2},
\ee
\be \label{eqN}
(Nr)'=1+2gr^2(1-\cR),
\ee
where now
\be \label{R}
\cR=\sqrt{1+ 2\frac{Nw'^2}{r^2} + \frac{(1-w^2)^2}{r^4}},
\ee
and primes denotes the derivatives with respect to r.
The third is a decoupled equation for $\sigma$
\be
(\ln\sigma)'=\frac{2g w'^2}{r\cR}
\ee
which can be easily solved once the YM function $w$ is found.

We are interested in the asymptotically flat configurations such that
the local mass function $m(r)$, defined through
\be
N=1-\frac{2m(r)}{r},
\ee
has a finite limit $m\to M$ as $r\to\infty$, while $\sigma\to 1$.
Like in the EYM case, it can be easily derived from the Eq. (\ref{eqN})
that finiteness of $M$ implies the following asymptotic behavior of the
YM function: $w=\pm 1+O(r^{-1})$. In this limit $\cR\to 1$ so the BI
non-linearity is negligible.

Let us first discuss the globally regular solutions which start at the origin
with the following series expansion (its convergence in a non-zero domain
may be proved by the standard methods \cite{BrFoMa94}):
\be \label{regw}
w=1-br^2+ \frac{br^4\left(3b(44b^2+3)+
4g[28b^2+3-(4b^2(48b^2+13)+3)\cR_0^{-1}])\right)}{30(4b^2+1)}+O(r^6),
\ee
\be \label{regN}
N=1+\frac{2r^2}{3} g\left(1-\cR_0\right)-\frac{16gb^2r^4}{15(4b^2+1)}
\left(g(\cR_0-1)^2+3b\cR_0\right)+O(r^6),
\ee
where $\cR_0$ is a limiting value of the square root at the origin
\be
\cR\to\cR_0=\sqrt{1+12 b^2}.
\ee
This local solution has a unique free parameter $b$.
In the limiting case $g=0$ it coincides with
that found in the flat space case \cite{GaKe99}, while to make contact with
the corresponding expansion of the EYM theory one has to take the limit
$g\to\infty$  with a simultaneous rescaling $b\to b/g$.
Near the origin the spacetime is flat.

Like in the EYM case, matching  of the local solution departing from the
origin as (\ref{regw},\ref{regN}) and meeting the conditions of asymptotic
flatness can be achieved for a discrete sequence of the free parameter
values $b=b_n(g)$ labeled by the number $n$ of zeroes of $w(r)$. The proof of
existence may be given along the lines of \cite{BrFoMa94,GaKe99},
here we do not enter into mathematical details concentrating rather
on the qualitative physical picture.

\begin{figure}
\unitlength1cm
\begin{picture}(14,11)
\put(1,0){\epsfig{file=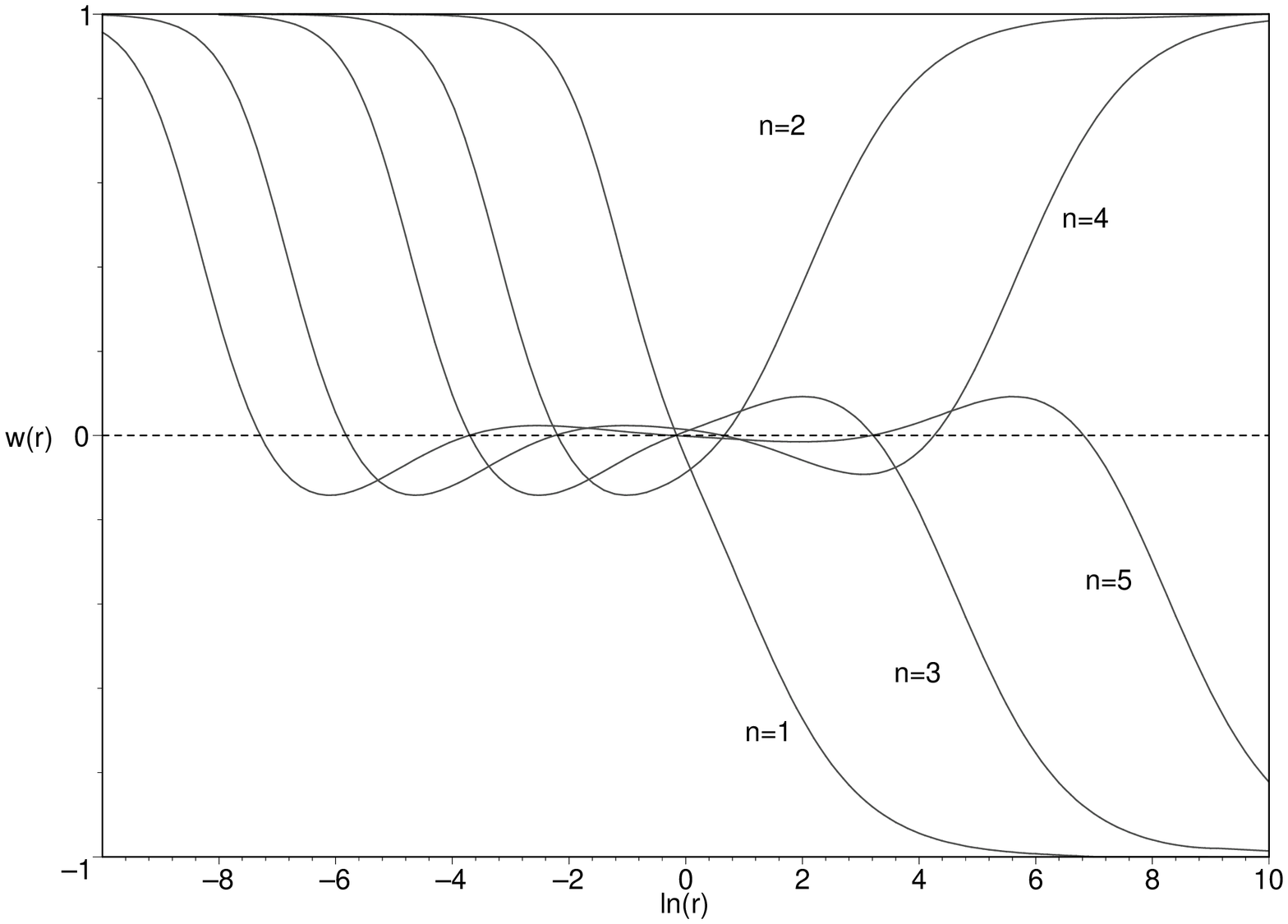,width=14cm,height=11cm}}
\end{picture}
\caption{First five solutions $w_n(r)$ for weak gravity ($g=.001$).
The oscillation region, which corresponds to the localization of an
unscreened magnetic charge inside the BIYM particle, expands with growing $n$.
The amplitude of the first and the last oscillation cycles remains roughly
the same, while the amplitude of the intermediate cycles tends to zero with
increasing $n$. All these curves practically coincide with those found
in the flat space BIYM theory by Gal'tsov and Kerner.}
\label{r001w}

\end{figure}

\begin{figure}
\unitlength1cm
\begin{picture}(14,11)
\put(1,0){\epsfig{file=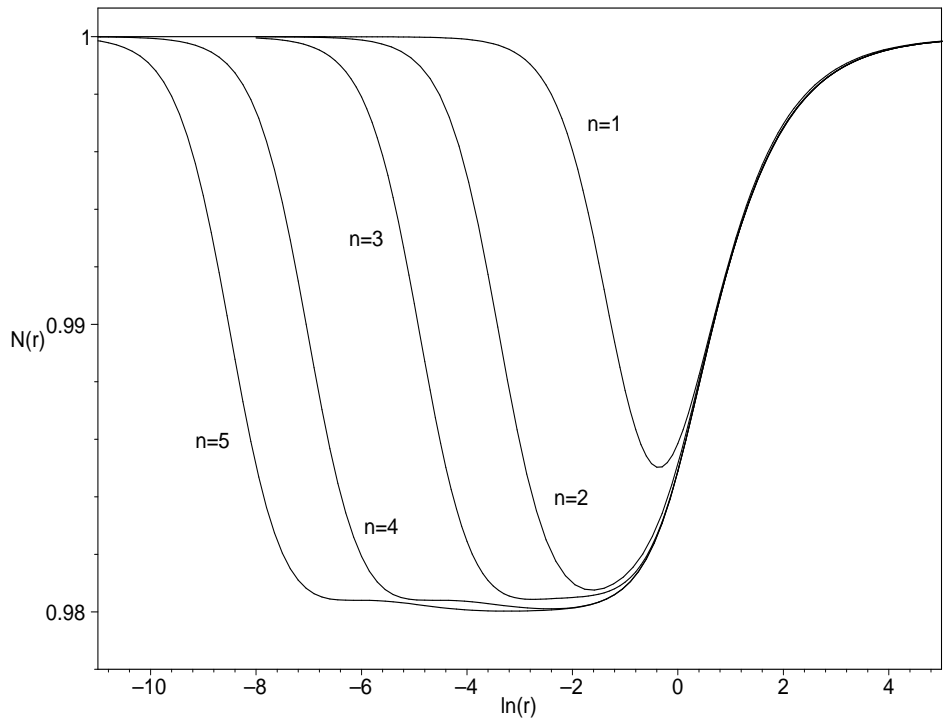,width=14cm,height=11cm}}
\end{picture}
\caption{Metric function  $N(r)$ for regular solutions
with $g=.001$. For all solutions the deviation of the metric from a flat one
is within $2\%$. For higher node numbers $n$ one observes a stabilization
of $N$ near the minimal value, this corresponds to the oscillation region
of $w(r)$.
}
\label{r001N}

\end{figure}

We have investigated numerical solutions in a large range of $g$.
Typical behavior for small $g$ is shown of Figs. (\ref{r001w},\ref{r001N}).
The YM curves for $g=.001$ are practically undistinguishable from those
found previously by Kerner and one of the authors \cite{GaKe99} in the
flat space BIYM model. The region of oscillations corresponds to an
unscreened Coulomb charge inside the particles. One can see that this
region expands both in the direction of small and large $r$ with growing
$n$. The metric deviate from the flat one rather weakly for small $g$
(weak gravity). One observes a stabilization of the metric in the region of
the $w$-oscillations.

The amplitude of oscillations of $w$ in the intermediate region decreases,
so in this region the solution can be regarded as approaching the embedded
Abelian solution $w\equiv 0$. As was observed by Wirschins, Sood and Kunz
\cite{WiSoKu00}, the metric for small $g$ also approaches the corresponding
Abelian Bion metric \cite{GiRa95} (with zero 'seed' mass in the singularity).
To avoid confusion it is worth noting
that for odd $n$ the Yang-Mills topology of the EBIYM regular solutions (kink)
is essentially different form that of the embedded Abelian one $w\equiv 0$
(trivial). Thus it would be misleading to say that the sequence of
non-Abelian EBIYM solutions converges to an Abelian one in the global sense.

\begin{figure}
\unitlength1cm
\begin{picture}(14,11)
\put(1,0){\epsfig{file=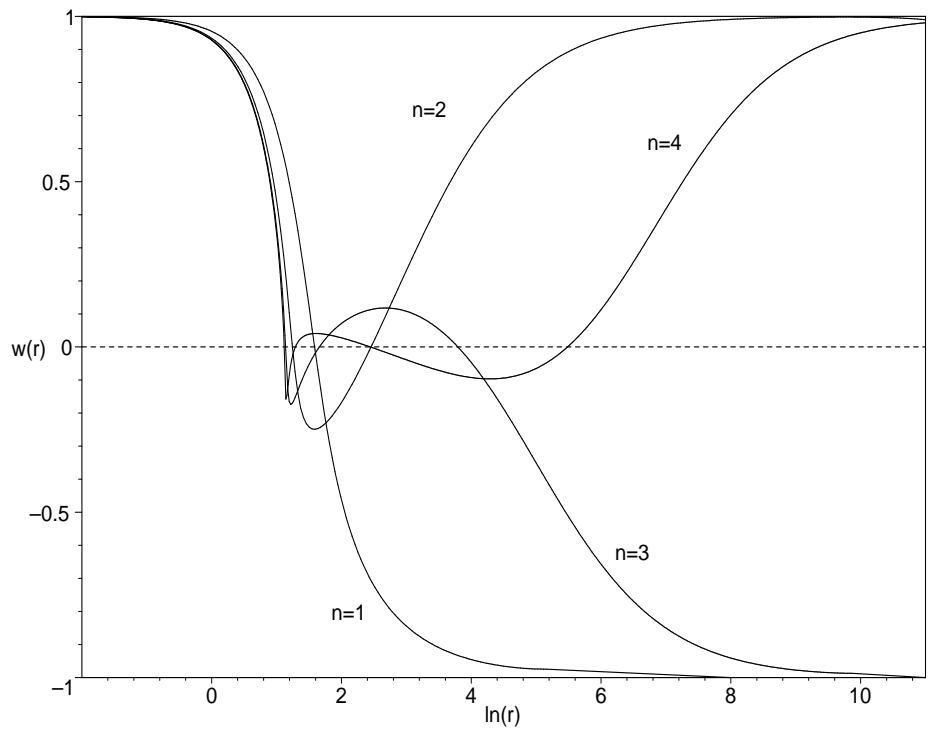,width=14cm,height=11cm}}
\end{picture}
\caption{Regular solutions $w_n(r)$ for strong gravity, $g=10$.
These curves are close to BK solutions in terms of a rescaled coordinate
$\tilde r =r/g^{1/2}$
.}
\label{r10w}

\end{figure}

For large $g$ the regular EBIYM solutions after the coordinate rescaling
$r\to \sqrt{g} r$ tend to the BK solutions of the EYM system,
see Fig.~\ref{r10w} for $g=10$.
Comparing this with the weak gravity behavior (Fig.~\ref{r001w})
we observe that gravity reduces the region of the unscreened charge.
The metric deviates substantially form the flat one. The function
$N(r)$ for large $n$ exhibits a deep well (an 'almost' horizon),
but remains always strictly positive.

The sequences $b_n$ of discrete values of the parameter in the expansion
(\ref{regw}) behaves very differently for small and large $g$. As was found
in \cite{GaKe99}, $b_n$ in the flat space are quite big with respect
to the corresponding BK values, and the sequence does not converge
with growing $n$. Contrary to this, the BK sequence $b_n$
rapidly converges to a limiting value $b_\infty$. In this case there is
an additional limiting solution with different space-time structure
\cite{BrFoMa94}.

\begin{figure}
\unitlength1cm
\begin{picture}(14,11)
\put(1,0){\epsfig{file=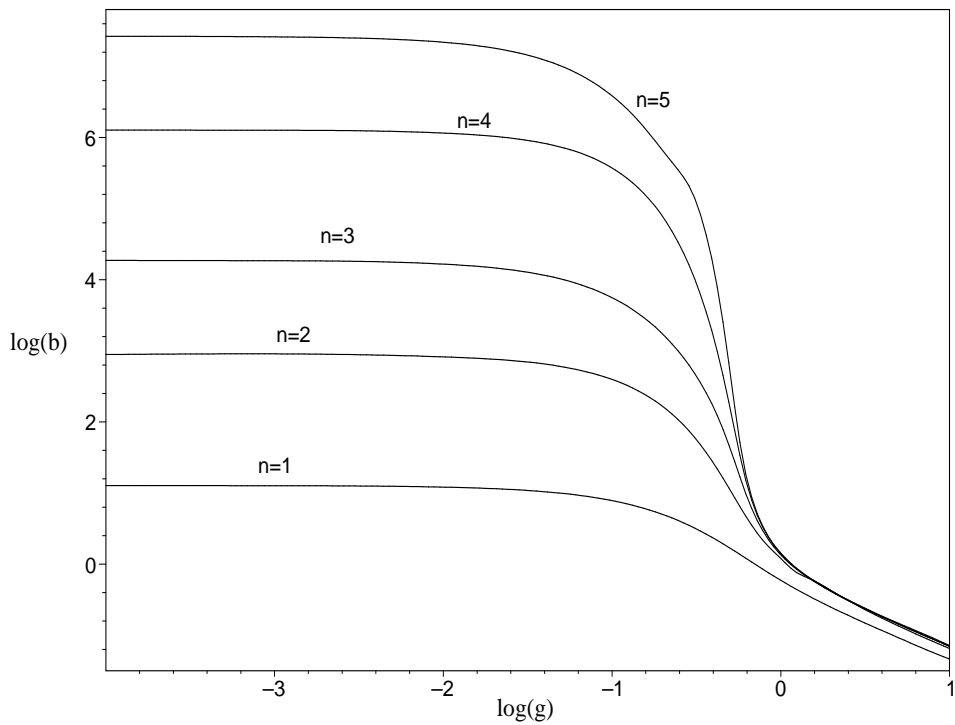,width=14cm,height=11cm}}
\end{picture}
\caption{Dependence of the parameters $b_n$ for regular solutions on the
effective gravity coupling constant $g$ (in logarithmic coordinates).
The left side of the picture correspond to the GK limit. With growing $g$
higher-$n$ curves merge, this is reminiscent of the
rapid convergence of the sequence $b^{BK}_n=\lim_{g\to\infty}b_n(g)/g$.
} \label{b}

\end{figure}

We have obtained
numerically the interpolating functions $b_n(g)$ for several $n$
clearly demonstrating  that
these two extrema are continuously related indeed (Fig.~\ref{b}).
For small values of $g$ these functions tend to the GK values \cite{GaKe99}
\be
b_n(0)=b_n^{GK},
\ee
while in the opposite limit
$g\to\infty$ one recovers the convergent sequence of rescaled
BK values \cite{BaMc88,VoGa98}
\be
b_n^{BK}=\lim_{g\to\infty} b_n(g)/g.
\ee
The corresponding solutions of the EYM system with the standard YM lagrangian
are recovered in terms of the rescaled coordinate $\tilde r =r/\sqrt{g}$.
In the intermediate region all parameter functions $b_n(g)$ were found
to be monotonously varying between the above extrema.

\begin{figure}
\unitlength1cm
\begin{picture}(14,11)
\put(1,0){\epsfig{file=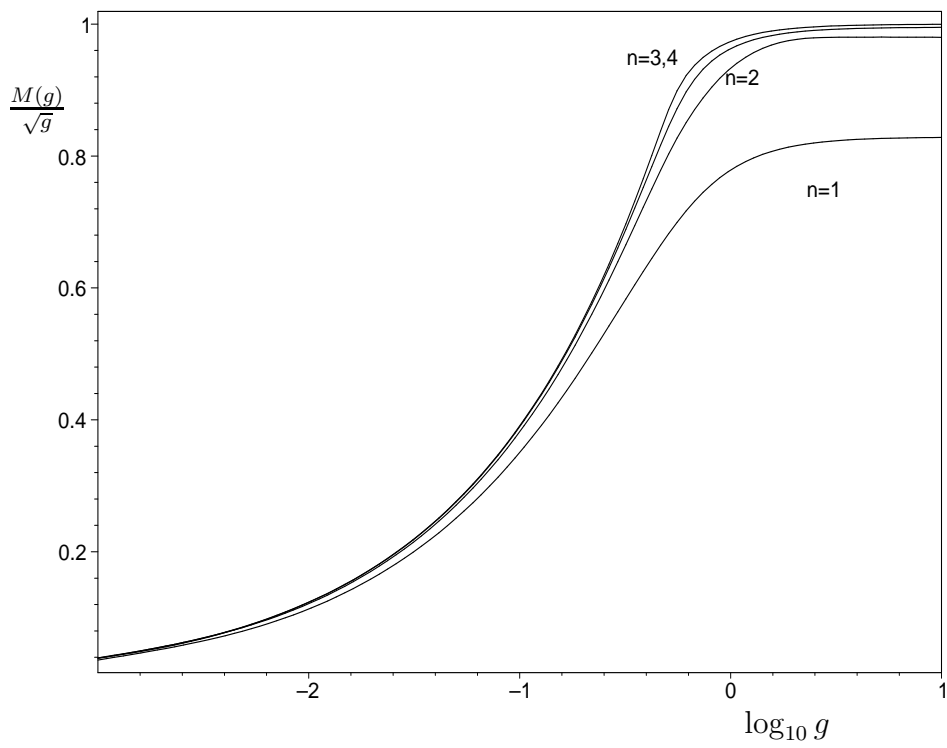,width=14cm,height=11cm}}
\put(1.2,8.5){$\frac{M(g)}{\sqrt{g}}$} \put(11,0.3){$\log_{10} g$}
\end{picture}
\caption{Dependence of masses of regular solutions on the gravity
coupling $g$}
\label{mass}

\end{figure}

Masses of the regular solutions as functions of the effective gravitational
coupling are shown on Fig.~\ref{mass}. For vanishing $g$ one recovers
the GK masses after a rescaling
\be
\lim_{g\to 0} M_n(g)\to M_n^{GK} g.
\ee
We recall \cite{GaKe99} that the sequence $M_n^{GK}$ converges to the
mass of an embedded Abelian solution. The reason is that the main
contribution to the mass comes from the region of oscillations where $w$
approaches an Abelian value $w\equiv 0$ with growing $n$. But for any
$n$ the limiting values $w(0), w(\infty)$ are equal to $\pm 1$, so, as
we have already noted, the global topology of non-Abelian solutions
is entirely different. In the limit of strong gravity one recovers the BK
masses after a rescaling
\be
\lim_{g\to \infty} M_n(g)\to M_n^{BK} \sqrt{g}.
\ee

Now discuss the black holes. These are parameterized by the horizon
radius $r_h$ and the value of the YM function $w_h=w(r_h)$ at the horizon.
The series expansions near the regular horizon reads:
\bea
w&=&w_h + \frac{w_h(w_h^2-1)}{r_hN'_h} (r-r_h)+ O\left((r-r_h)^2\right) \nonumber\\
N&=&N'_h (r-r_h)+ O\left((r-r_h)^2\right),\\
N'_h&=&\frac{1}{r_h}\left[1+2g^2 r_h^2
\left(1+\sqrt{1+(w_h^2-1)^2/r_h^4}\right)\right]. \nonumber
\eea
Asymptotically flat solutions with such boundary condition are likely
to exist for any horizon radii $r_h$. Exterior black hole solutions are
very similar to regular solutions, especially for small $r_h$. So our main
interest is in the interior solutions. We start by listing
various series solutions that can be obtained near the singularity.

Let us first explore the series expansion for an
embedded Abelian solution \cite{GiRa95,Ra97}. For the unit
magnetic charge (what corresponds to $w\equiv 0$) the local mass function
satisfies the equation
\be
m'=g\left(\sqrt{r^4+1}-r^2\right).
\ee
Expanding the square root at small $r$ and integrating one obtains:
\be \label{eab}
N=-\frac{2m_0}{r}+1-2g+\frac23 gr^2-\frac15 gr^4+O(r^8).
\ee
The 'seed' mass parameter $m_0$ may be positive, negative or zero. Positive
$m_0$ corresponds to a timelike singularity of the Schwarzschild type.
Negative $m_0$ corresponds to a spacelike singularity, in which case an
internal horizon also exist (though contrary to a more common example of the
Reissner-Nordstr\"om metric, the 'local charge' term $Q^2/r^2$
now is absent). Vanishing $m_0$  is particularly interesting. The metric
near the origin is then locally flat unless $g=1/2$ in which case
the limiting value $N(0)$ shrinks to zero. For other values of $g$
there is a conical
singularity \cite{GiRa95}, the critical value $g=1/2$ corresponding to
an extremal deficit angle. Within the present framework the embedded
Abelian black holes correspond to an identically vanishing function $w$.
It can be shown that the local series solution starting at $r=0$ with
zero initial value $w_0=0$ and arbitrary $m_0$ generates this global
embedded Abelian solution.

For non-Abelian solutions the value of the YM function at the singularity
$w(0)=w_0$ should therefore be non-zero.
We have found the generalization of the series expansion (\ref{eab})
with $w_0\neq 0$. It is valid for the non-zero seed mass parameter $m_0$:
\be \label{qabw}
w=w_0\left(1+\frac{er}{2m_0}+\frac{er^2}{16m_0^2}\left[3(2e-1)
-4ge\right]\right) +O(r^3),
\ee
\[e
N=-\frac{2m_0}{r} +1-2ge+\frac{3gew_0^2 r}{2m_0}+
gr^2\left(\frac23+\frac{ew_0^2}{12 m_0^2}\left[3(5e-2)-8ge\right]\right)
+ O(r^3),
\]
where $e=1-w_0^2$. This local solution has two free parameters $w_0, m_0$.
For $w_0=0$ it coincides with (\ref{eab}). If $w_0\neq 0$  becomes singular
in the limit $m_0=0$. The search for local solutions with $m_0=0$ shows that
in this case either $w_0=0$ in which case we come back to the Abelian
embedded solution $w\equiv 0$, or $w_0=\pm 1$, then we recover the regular
solution (\ref{regw},\ref{regN}). It is also worth noting that, contrary
to the EYM case, in the EBIYM theory there are no local solutions with
$N\sim r^{-2}$ behavior at the singularity (Reissner-Nordstr\"om type).

Since the asymptotic solution (\ref{qabw}) fails to contain a sufficient
number of free parameters to be a generic solution, (this number is equal
to three for the system of equations (\ref{eqw},\ref{eqN})), the question
is how a generic solution looks like near
the singularity, in particular, whether it admits any series expansion.
In the case of the ordinary Yang-Mills lagrangian such an expandable solution
does not exist at all, one finds that a generic solution has a non-analytic
oscillating behavior \cite{DoGaZo96}. Here the situation is different,
although the generic local solution around the singularity still exhibits
non-analyticity in terms of the variable $r$. It turns out to be series
expandable but in terms of the $\sqrt{r}$:
\bea \label{gen}
w&=&w_0+a\sqrt{r}-\frac{a^2w_0r}{e}+
\frac{a\left( 24 g^2 a^2-32 g^2 a^2 e-3c^2\right)r^{3/2}}{32 g^2 e^2}-\nonumber\\
&&\frac{a^2\left (\left (16 g^2a^2-
32 g^2a^2e-15 c^2\right)w_0-8 g^2 cae\right) r^2}{32 g^2 e^3}+O(r^{5/2}),
\nonumber\\
N&=&1-2\frac{e^2}{a^2r}+c\sqrt{r}-\frac{a\left(3 w_0
c+ag^2e\right)r}{e}+O(r^{3/2}).
\eea

This local solution contains three free parameters $w_0, a, c$.
The singularity is of the Schwarzschild type, the seed mass $m_0$ is strictly
positive. Two leading terms in expansion of 'kinetic'
(negative for $N<0$) and 'potential' (positive) terms in $\cR$ cancel,
so the leading behavior in singularity is
\be
\cR\sim \frac{{\rm 3c}}{4gr^{3/2}}.
\ee

\begin{figure}
\unitlength1cm
\begin{picture}(14,11)
\put(1,0){\epsfig{file=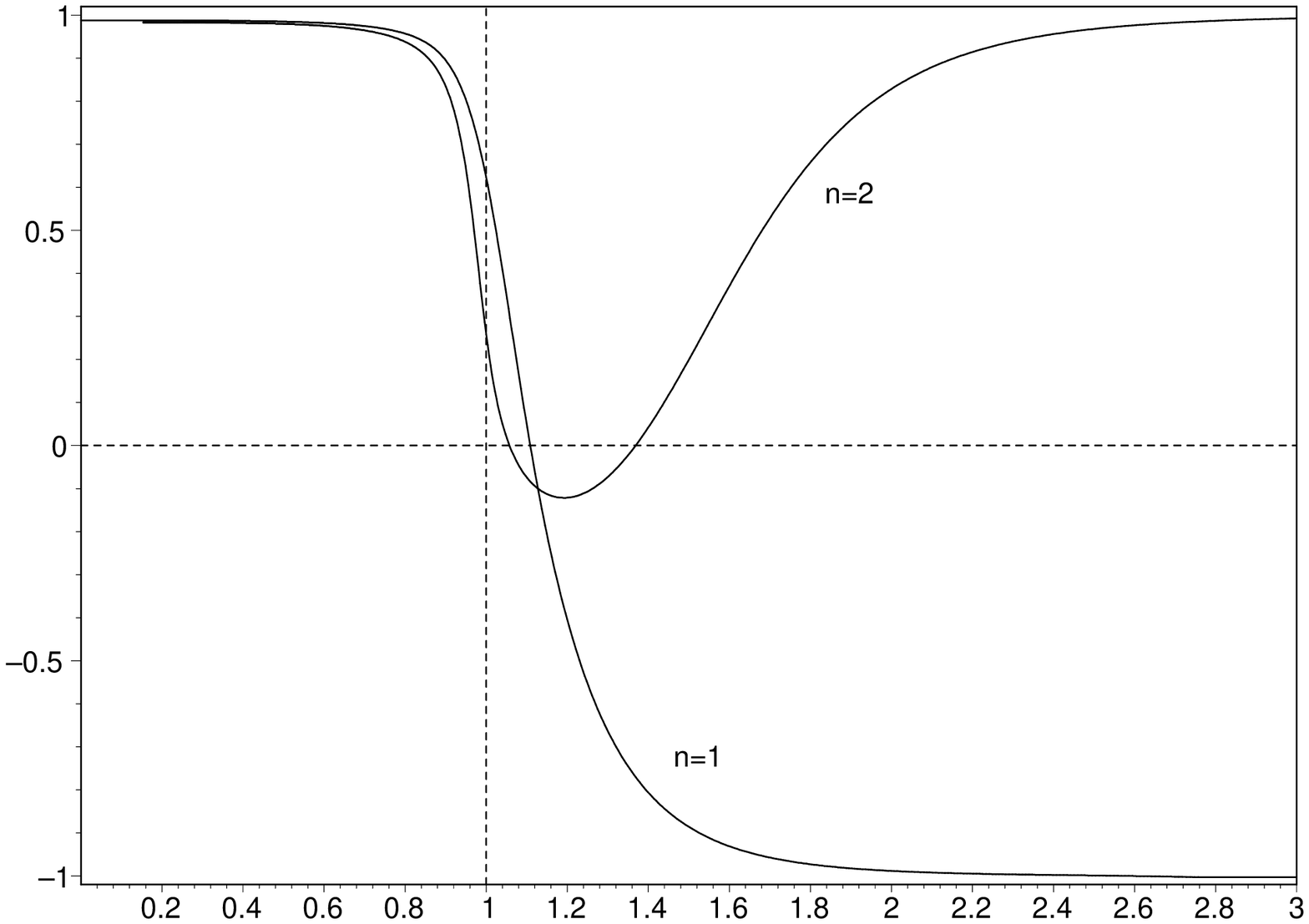,width=14cm,height=11cm}}
\put(1.2,8.5){$w(r)$} \put(4.8,1.8){horizon}
\put(11,0.3){$r^{1/8}$}
\end{picture}
\caption{YM function $w(r)$ for $n=1,2$ EBIYM black holes with
$r_h=1.0,\; g=1.0,$. At the singularity these curves tend to constant
values non-equal to unity.}
\label{bhw}
\end{figure}

\begin{figure}
\unitlength1cm
\begin{picture}(14,11)
\put(1,0){\epsfig{file=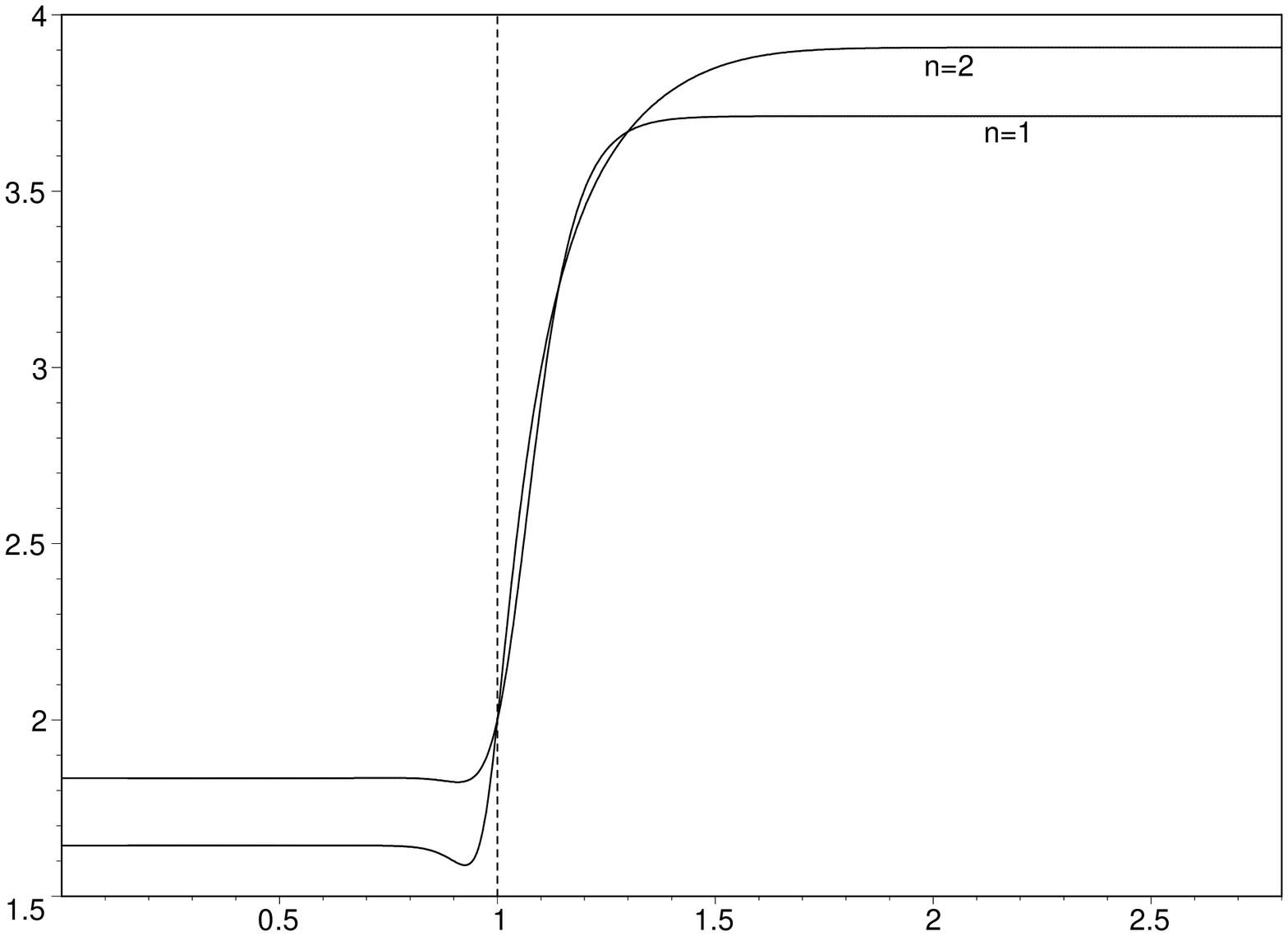,width=14cm,height=11cm}}
\put(1.2,8.5){$m(r)$} \put(4.8,8.8){horizon}
 \put(11,0.3){$r^{1/8}$}
\end{picture}
\caption{The mass function $m(r)$ function for the EBIYM black holes with
$r_h=1.0,\; g=1.0,\; n=1,2$. Inside the horizon $m(r)$ remains a smooth
function tending to a constant limit at the singularity. There are no
internal horizons, the singularity is timelike.}
\label{bhm}
\end{figure}

\begin{figure}
\unitlength1cm
\begin{picture}(14,11)
\put(1,0){\epsfig{file=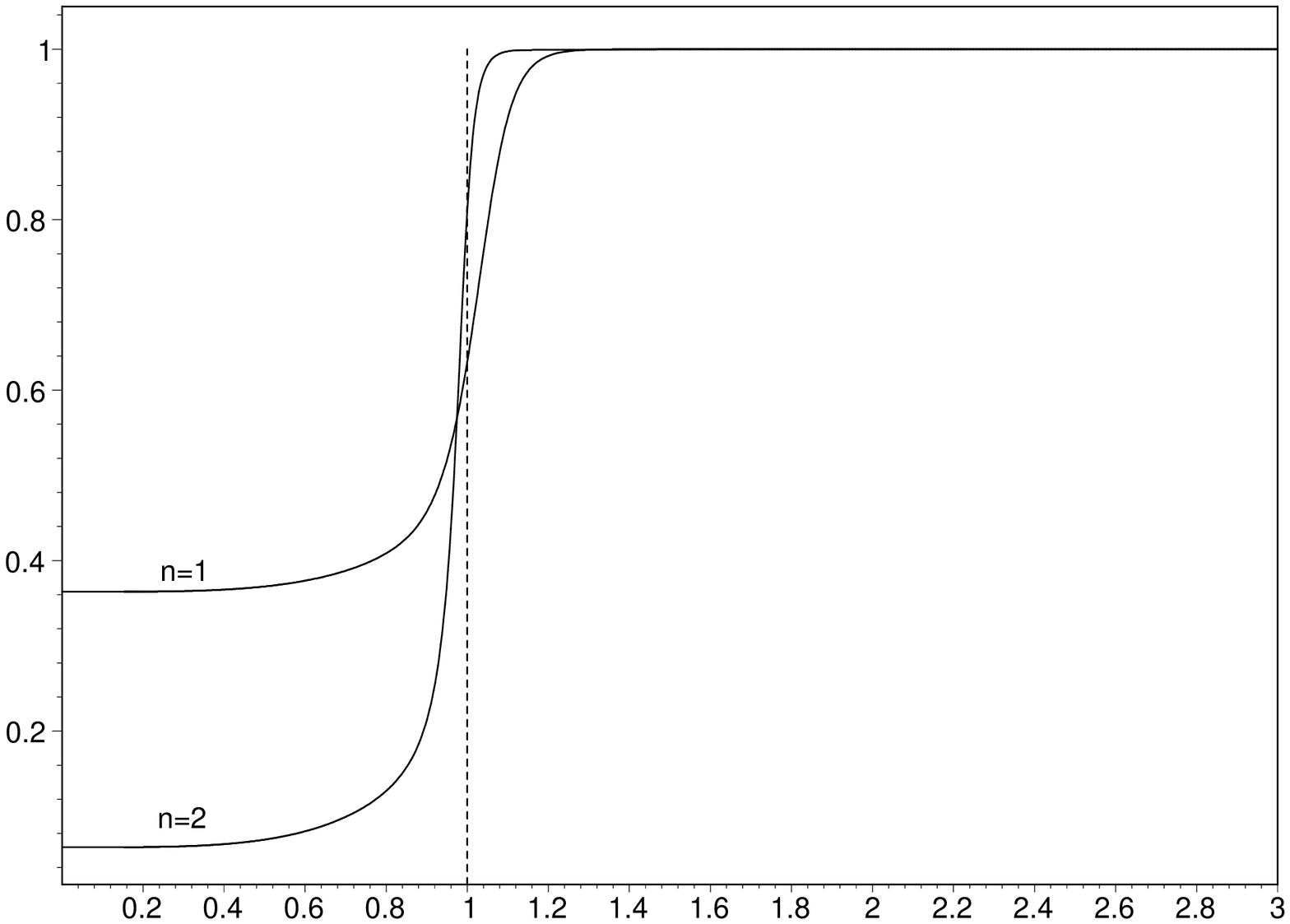,width=14cm,height=11cm}}
\put(1.2,8.5){$\sigma(r)$} \put(6.3,2.0){horizon}
\put(11,0.3){$r^{1/8}$}

\end{picture}
\caption{The metric function $\sigma(r)$ for the EBIYM black holes with
$r_h=1.0,\; g=1.0,\; n=1,2$.
}
\label{bhs}
\end{figure}

We have tested for various $r_h, g, n$ that a continuation of the exterior
black hole solutions under the horizon meets this generic asymptotic
solution indeed. In all numerical experiments the function $N$ remained
negative under the horizon and no internal Cauchy horizons were met.
Typical global black hole solutions are shown on the
Figs.~\ref{bhw},\ref{bhm},\ref{bhs} for $r_h=1, g=1, n=1,2$.
The YM function $w$ outside the horizon has qualitatively the same
behavior as in the regular case. Inside the horizon it remains perfectly
smooth and tends to a finite limit $w_0\neq 0$ at the singularity.
Recall that for the ordinary quadratic YM lagrangian the function $w$
inside the horizon of the EYM black holes has a rather sophysticated
behavior: while $w$ itself tends to a finite limit
$w_0$ as well, its derivative exhibits a sequence of infinitely increasing
absolute values at tiny intervals, whose length tends to zero
\cite{DoGaZo96,GaDoZo97}. Such a behavior causes oscillations of the local
mass $m(r)$ with an infinitely increasing amplitude. At the beginning of
each oscillation cycle the metric function $N(r)$ takes values very close
to zero ('almost' Cauchy horizons), then an exponential growth of $m(r)$
starts. After a few oscillation cycles the maximal values of $m$ attained
in subsequent cycles become of the googolplexus order,
obviously lying beyond any classical bounds.
Contrary to this, in the EBIYM case we observe a smooth $m(r)$
inside the horizon up to the singularity where $m(r)$ has a finite positive
value (Fig.~\ref{bhm}). Similarly, the second metric function $\sigma$
tends smoothly to a finite value at the singularity (Fig.~\ref{bhs})

We conclude with the following remarks.
The BK solutions were found for the gravity coupled Yang-Mills theory with
the usual quadratic lagrangian. Now they were shown to be a strong
gravity limit of the gravity coupled BIYM theory which can be interpreted
as an effective YM theory including string quantum corrections. Within
this theory there exists a limit of decoupled gravity, in which
qualitatively similar solutions continue to exist. This shows the way
how the BK solutions can be incorporated into the (quantum) string theory.

A particularly interesting implication of this reasoning is the resolution
of the problem of `violent' oscillating singularities typical for
hairy EYM black holes. Born-Infeld corrections perfectly regularize
the behavior of the metric near the singularity. It is worth noting
that the generic singularity is timelike, in conformity with the
strong cosmic censorship hypothesis. In numerical experiments we did not
observe internal horizons. In principle, such horizons could emerge in pairs,
this question is worth to be investigated in more detail.

This work was supported in part by the RFBR grant 00-02-16306.

\newpage

\begin{figure}
\unitlength1cm
\begin{picture}(14,11)
\put(1,0){\epsfig{file=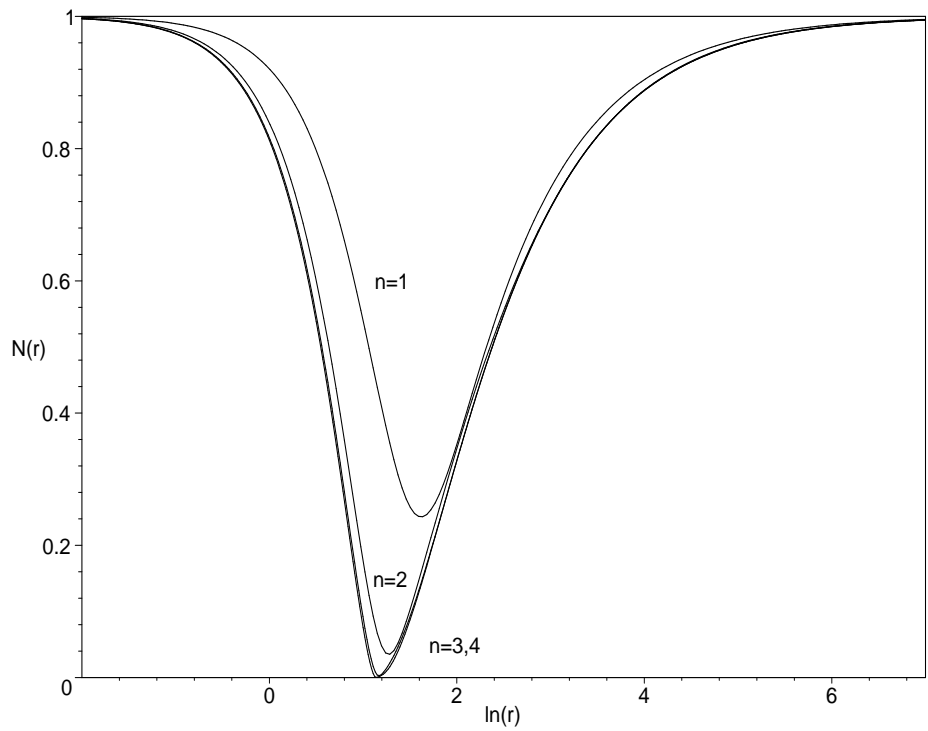,width=14cm,height=11cm}}
\end{picture}
\caption{Metric functions $N_n(r)$ for regular solutions with $g=10$.
For higher $n$ the curves come very close to but never reach zero:
$N_3^{min}\approx .003,\, N_4^{min}\approx .0003$.
}
\label{r10N}

\end{figure}


\begin{thebibliography}{99}

\bibitem{BaMc88}
R.~Bartnik and J.~McKinnon,
Phys. Rev. Lett. 61 (1988) 141.


\bibitem{VoGa98}
M.S. Volkov and D.V. Gal'tsov,
Phys. Reports C 319 (1999) 1, hep-th/9810070.

\bibitem{GaVo91}
D.V. Gal'tsov and M.S. Volkov,
Phys. Lett.  B 273 (1991) 255.

\bibitem{VoGa89}
 M.~S. Volkov, and D.~V. Galt'sov,
 JETP Lett. 50 (1989) 346; Sov. J. Nucl. Phys. 51 (1990) 747.

\bibitem{Bi90}  P. Bizon,
 Phys. Rev. Lett. 64 (1990) 2844;

\bibitem{Ku90}
H.~P. Kunzle, and A.~K.~M. Masood-ul-Alam,
 J. Math. Phys. 31 (1990) 928.

\bibitem{DoGaZo96}
E.E. Donets, D.V. Gal'tsov, and M.Yu. Zotov
Phys. Rev. D 56 (1997) 3459.

\bibitem{GaDoZo97}
D.V. Gal'tsov, E.E. Donets,  and M.Yu. Zotov
JETP Lett. 65 (1997) 895.

\bibitem{BrLaMa98}
P. Breitenlohner, G. Lavrelashvili, and D. Maison,
Nucl. Phys. B 524 (1998) 427.

\bibitem{Po98}
J. Polchinski, String Theory, Vols. I and II, CUP, 1998.

\bibitem{Ts97}
A.Tseytlin,
Nucl. Phys. B 501 (1997) 41.

\bibitem{GaGoTo98}
J.P.~Gauntlett, J.~Gomis and P.K.~Townsend,
JHEP 01 (1998) 003 .

\bibitem{BrPe98}
D.~Brecher and M.J.~Perry,
 Nucl. Phys. B 527 (1998) 121.

\bibitem{GrMoSc99}
N.~Grandi, E.F.~Moreno and F.A.~Shaposhnik,
hep-th/9901073.

\bibitem{Pa99}
 J.H. Park, Phys. Lett. B458 (1999) 471.

\bibitem{Gr99}
 N. Grandi, R.L. Pakman, F.A. Schaposnik, and G. Silva,
 Phys. Rev. D 60 (1999) 125002.

\bibitem{GaKe99}
 D. Gal'tsov, and R. Kerner,
Phys. Rev. Lett. (to appear); hep-th/9910171.

\bibitem{WiSoKu00}
M. Wirschins, A. Sood, and J. Kunz,
hep-th/0004130.

\bibitem{BrFoMa94}
P.~Breitenlohner, P.~Forgacs, and D.~Maison,
Comm. Math. Phys.  163 (1994) 141.

\bibitem{GiRa95}
G.W. Gibbons and D.A. Rasheed,
Nucl. Phys. B 454 (1995) 185.

\bibitem{Ra97}
 D.A. Rasheed,
 hep-th/9702087.
\end{thebibliography}
\end{document}